\documentclass[aps,prl,preprint,nopacs,superscriptaddress]{revtex4-1}

\usepackage{lineno}
\usepackage{graphicx}
\usepackage{verbatim}
\usepackage{mathrsfs}
\usepackage{gensymb}
\usepackage{color}
\usepackage{ulem}

\usepackage{amsmath,amsfonts,amssymb}
\usepackage{graphicx}

\newcommand{\bee}{\begin{equation}}
\newcommand{\ee}{\end{equation}}
\def\3{2.8in}    
\def\2{2.5in}
\def\4{3.0in}
\def \beq {\begin{equation}}
\def \eeq {\end{equation}}
\newcommand{\Z}{\mathbb{Z} }   
\newcommand{\kp}{{$k\cdot p$ }}
\newcommand{\ie}{{\it i.e. }}

\newcommand{\K}{\mathcal{K}}
\newcommand{\T}{\mathcal{T}}
\newcommand{\I}{\mathcal{I}}
\newcommand{\Q}{\mathcal{Q}}

\newcommand{\Mz}{\mathcal{M}_z}

\newcommand{\Gy}{\mathcal{G}_y}
\newcommand{\Gx}{\mathcal{G}_x}

\newcommand{\Sy}{\mathcal{S}_y}
\newcommand{\Sx}{\mathcal{S}_x}
\newcommand{\ra}{\textbf{r}_1}

\begin{document}

\title{Glide Symmetry Protected Higher-Order Topological Insulators from Semimetals with butterfly-like Nodal Lines}

\author{Xiaoting Zhou\footnote{physxtzhou@gmail.com}}
\affiliation{Department of Physics and Astronomy, California State University, Northridge, CA 91330, USA}

\author{Chuang-Han Hsu}
\affiliation{Department of Electrical and Computer Engineering, Faculty of Engineering, National University of Singapore, Singapore 117583}
\affiliation{Institute of Physics, Academia Sinica, Taipei 11529, Taiwan, R.O.C}

\author{Cheng-Yi Huang}
\affiliation{Institute of Physics, Academia Sinica, Taipei 11529, Taiwan, R.O.C}
\affiliation{Department of Physics and Astronomy, California State University, Northridge, CA 91330, USA}

\author{Mikel Iraola}
\affiliation{Donostia International Physics Center, 20018 Donostia-San Sebastian, Spain}
\affiliation{Department of Condensed Matter Physics, University of the Basque Country UPV/EHU, Apartado 644, 48080 Bilbao, Spain}

\author{Juan L. Ma\~nes}
\affiliation{Department of Condensed Matter Physics, University of the Basque Country UPV/EHU, Apartado 644, 48080 Bilbao, Spain}

\author{Maia G. Vergniory}
\affiliation{Donostia International Physics Center, 20018 Donostia-San Sebastian, Spain}
\affiliation{IKERBASQUE, Basque Foundation for Science, Maria Diaz de Haro 3, 48013 Bilbao, Spain}

\author{Hsin Lin}
\affiliation{Institute of Physics, Academia Sinica, Taipei 11529, Taiwan, R.O.C}

\author{Nicholas Kioussis\footnote{nick.kioussis@csun.edu}}
\affiliation{Department of Physics and Astronomy, California State University, Northridge, CA 91330, USA}

\pacs{}


\begin{abstract}
Most topological insulators discovered today in spinful systems can be transformed from topological semimetals (TSMs) with vanishing bulk gap via introducing the spin-orbit coupling (SOC), which manifests the intrinsic links between the gapped TI phases and the gapless TSMs. Recently, we have proposed a new family of TSMs in time-reversal invariant {\it spinless} systems, which host butterfly-like nodal-lines (NLs) consisting of a pair of identical concentric intersecting coplanar ellipses (CICE). In this Communication, we unveil the intrinsic link between this exotic class of nodal-line semimetals (NLSMs) and a $\Z_{4}$ = 2 topological crystalline insulator (TCI), by including substantial SOC. We demonstrate that in three space groups ({\it i.e.} $Pbam$ (No.55), $P4/mbm$ (No.127) and $P4_2/mbc$ (No.135)), the TCI supports a fourfold Dirac fermion on the (001) surface protected by two glide symmetries, which originates from the intertwined drumhead surface states of the CICE NLs. The higher order topology is further demonstrated by the emergence of one-dimensional helical hinge states, indicating a new higher order topological insulator protected by a glide symmetry.
\end{abstract}

\maketitle

The discovery of the quantum spin Hall effect (QSHE)~\cite{kane2005,kane2005a,bernevig2006} and topological insulators (TIs)~\cite{Fu2007prl,Moore2007,roy2009} which are protected by time-reversal symmetry (TRS), has indicated that symmetry plays a crucial role in classifying the topology of free fermion states~\cite{Altland1997,kitaev2009,Ryu2010}. Subsequently, the concept has been generalized to spatial symmetries in crystalline systems. 
For instance, topological crystalline insulators (TCIs)~\cite{Fu2011} are protected by other space-group symmetries ($\Q$)~\cite{Slager2013,Sato2014}, such as mirror ~\cite{Hsieh2012} and rotational symmetries~\cite{Song2017,Fang2019,Zhou2018a}. Consequently, such type of systems are known to harbor symmetry-protected topological (SPT) phase.~\cite{senthil2015}. 

In free fermion systems, SPT insulators harbor a central paradigm referred to as the bulk-boundary correspondence~\cite{jackiw1976}. A $d$-dimensional bulk with gapped excitations hosts anomalous gapless, topologically nontrivial boundary states in lower $(d-1)$ dimensions~\cite{Fu2007prl,Qi2011}. 
Recently, a higher-order bulk-boundary correspondence has been uncovered in types of TCI, which exhibit a gapped $(d-1)$-dimensional boundary but a gapless $(d-2)$-dimensional boundary~\cite{Benalcazar2017a,Benalcazar2017,Song2017,Schindler2017a}. Hence, this type of TCIs are dubbed higher-order topological insulators (HOTIs)~\cite{Benalcazar2017a,Benalcazar2017,Song2017,Schindler2017a,Schindler2018a,BernevigPRL2019,Park2019a,ZhangPRL2020,RenPRL2020}. In general, an $n$th order topological insulator describes a TCI in $d$-dimensions having symmetry protected $(d-n)$-dimensional gapless boundary states, but gapped otherwise, when the sample geometry is properly selected, being compatible with $\Q$. HOTIs protected by various symmetries have been studied, such as $\mathcal{C}^z_4\T$, the mirror symmetry~\cite{Schindler2017a}, and the inversion symmetry~\cite{Schindler2018a}, respectively.

In contrast to the gapped topological phase, a topological semimetal (TSM) has gapless bulk band structures, which are characterized by the topologically robust band-crossings manifolds between occupied and unoccupied bands in momentum space. Among them, nodal-line semimetals (NLSMs)~\cite{Burkov2011a,Fang2016}, which harbor one-dimensional (1D) nodal lines (NLs), possess the highest variability. NLSMs with NLs integrated in various configurations have been studied under the assumption that the spin-orbit coupling (SOC) is negligible or absent, \textit{e.g.,} a chain link~\cite{Bzdusek2016,Chang2017b,Yan2017}, a Hopf link~\cite{Chang2017b}, and a knot~\cite{Bi2017}. 

However, from another perspective, it is intuitive to raise the question whether additional topology could be unearthed when these intricate degenerate links are gapped out by substantial SOC. The intrinsic link between gapped TIs or TCIs and gapless TSMs is essential to trace the origin of the topology of the insulating phase due to the band inversions and the evolution of the surface states. It has been known that a $\Z_{2}$ strong TI can be realized from NLSMs with a single nodal ring when SOC is included~\cite{Rappe2015prl,Duan2016prb,Weng2017prb}, and a HOTI from NLSMs with monopole nodal lines~\cite{BernevigPRL2019}, which represent the intrinsic link between a gapped topological phase and a gapless TSM. Nevertheless, similar studies for NLSMs with complex NL configurations remain deficient. In this work, we unveil the intrinsic link between an exotic class of nodal-line semimetals (NLSMs) and a $\Z_{4}$ = 2 topological crystalline insulator (TCI). The new type of NLSM has been proposed in Ref.~\cite{Xiaoting2020a} recently in {\it spinless} systems, which hosts a butterfly-like NL consisting of a pair of concentric intersecting coplanar ellipses (CICE) residing on a plane in $k$ space as shown in Fig.\ref{Fig1}a. 


In this Communication, we include the $Pbam$ (No.55) symmetry-invariant SOC in a minimal tight-binding (TB) model which exhibits CICE in Ref.~\cite{Xiaoting2020a}, and (i) demonstrate that the CICE act as the origin of a TCI protected by two glide symmetries. With substantial SOC, as shown in Fig.\ref{Fig1}b, the CICE become anticrossing, thus driving a phase transition from NLSM to a TCI with $\Z_{4} = 2$~\cite{Po2017,Kruthoff2017,Khalaf2017a,Song2018b} 
due to the fact that the CICE are essentially sets of NLs stemming from the double-band-inversion (DBI). 
(ii) Consequently, the intertwined drumhead surface states (DSSs) on the (001) surface (wallpaper group (WG) $pgg$) stemming from the CICE nodal lines, evolve to the topological surface states (TSSs) with a fourfold Dirac fermion~\cite{Wieder2018a,young2015} protected by the two glide symmetries, corresponding to the TCI as shown in Fig.\ref{Fig1}b and Fig.\ref{Fig2}d, respectively. (iii) We further uncover the higher-order topology of the system featuring the 1D helical hinge states when the sample geometries are distinctively and properly selected. Moreover, it is the new HOTI protected by a nonsymmorphic glide symmetry.

\section{Results}

\begin{figure}
\begin{centering}
\includegraphics[width=\linewidth]{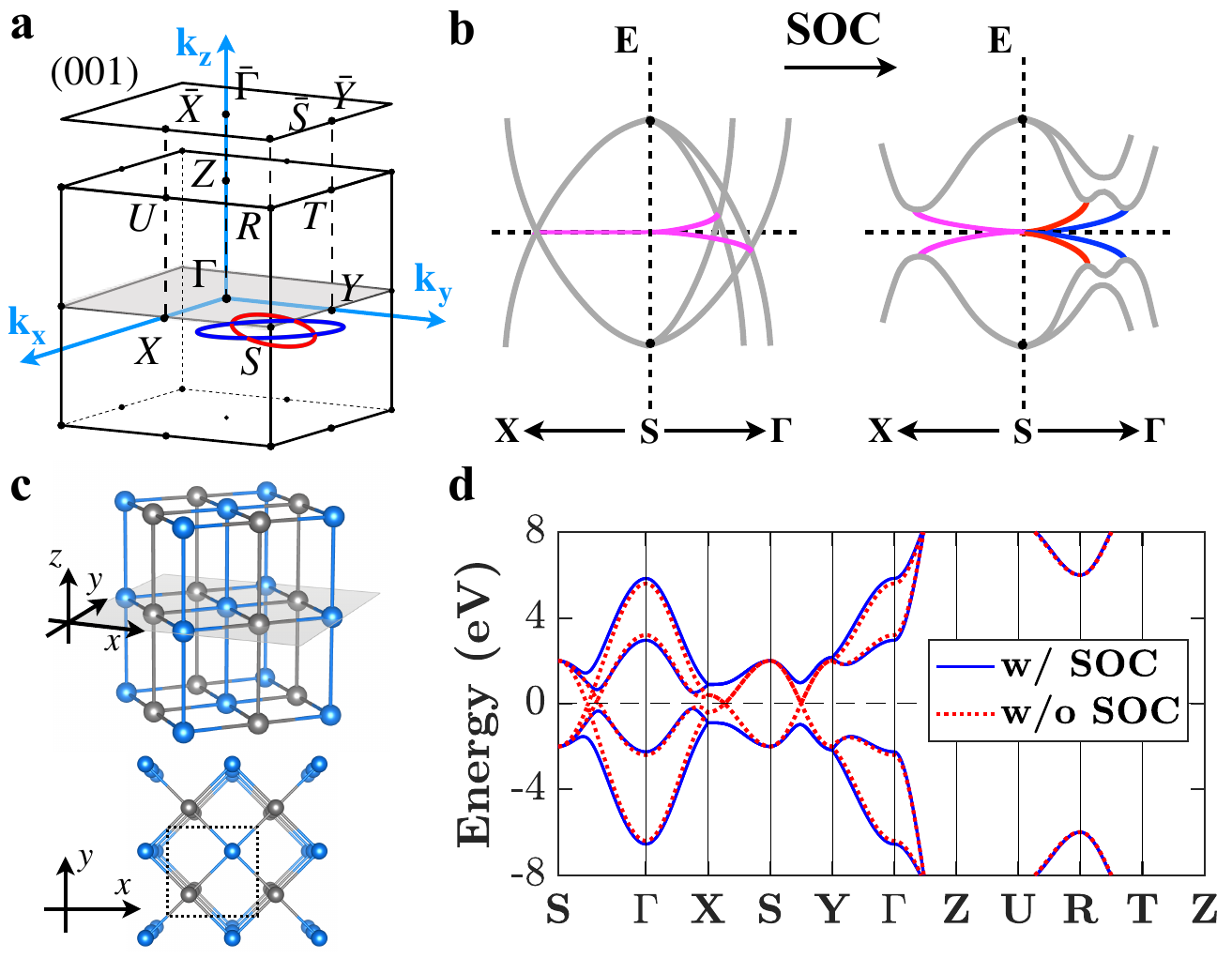}
\par\end{centering}
\centering{}
\caption{
{\bf Schematic phase transition from a NLSM to a TCI, crystal structure and band structures.} {\bf a,} Brillouin zone (BZ) of bulk and the projected $(001)$ surface in SG $Pbam$ (No. 55), with the high-symmetry points. The CICE nodal lines are on the $xy$-plane centred at $S$ point. {\bf b,} Schematic band structures demonstrating the SOC-driven transition from a NLSM exhibiting the CICE NLs~\cite{Xiaoting2020a} (left panel) to a TCI (right panel). Gray denotes the bulk states, and the magenta, red and blue indicate the (001) surface states, \ie the interwined DSSs of the NLSM (left panel), and the TSSs of the TCI with two intertwined Dirac cones (right panel).
{\bf c,} Orthorhombic crystal structure of the lattice model in Eq.(\ref{H0}) and (\ref{HSOC}), consisting of a lattice with two sublattices A (in blue) and B (in gray). {\bf d,} Band structure of the model with and without SOC, respectively.} 
\label{Fig1}
\end{figure}

\textbf{The lattice model.} The CICE can be sustained by two glide mirror symmetries and only nine space groups (SGs) are feasible to host it~\cite{Xiaoting2020a} . The minimal 4-band TB model for the \textcolor{black}{spinless} CICE NLSM in SG $Pbam$ (No. 55) is a lattice, consisting of two sublattices denoted by A (gray) and B (blue) which occupy the 2a Wyckoff position at $\mathbf{r}_A=(0, 0, 0)$ and $\mathbf{r}_B=(\frac{1}{2}, \frac{1}{2}, 0)$ in the unit cell (see Fig.~\ref{Fig1} c for the structure). 
There are two orbitals, $p_z$ and $d_{xy}$, for each sublattice, described by the Pauli matrix $\bm{\sigma}$ and $\bm{\tau}$ for the A and B sublattices, respectively. $\sigma_0$ and $\tau_0$ are identity matrices. 
The SG $Pbam$ (No.55) can be generated by the mirror symmetry, $\Mz=\lbrace m_{001}\vert 000 \rbrace$, and the two glide-mirror symmetries, $\Gx=\lbrace m_{100}\vert\frac{1}{2}\frac{1}{2}0 \rbrace$ and $\Gy=\lbrace m_{010}\vert\frac{1}{2}\frac{1}{2}0 \rbrace$, normal to the [100] and [010] directions, respectively, accompanied by a translation of $[\frac{1}{2}\frac{1}{2}0]$.
For a {\it spinless} system, employing the basis $\Psi = (p_{z}^A, d_{xy}^A, p_{z}^B, d_{xy}^B)^T$, 
the symmetry-constrained TB Hamiltonian is of the form, 
\begin{eqnarray}\label{H0}
H_0(\mathbf{k}) &=& [(\alpha \textrm{cos}k_x +\beta \textrm{cos}k_y + \gamma \textrm{cos}k_z) +\delta_0]\tau_{0} \sigma_{3} \\
& + & \textrm{cos}\frac{k_x}{2} \textrm{cos}\frac{k_y}{2} \textrm{cos}{k_z} (\lambda_{10}\tau_{1} \sigma_{0} + \lambda_{13} \tau_{1} \sigma_{3})\nonumber \\
& + & \textrm{sin}{k_z} (\lambda_{32} \tau_{3} \sigma_{2}) 
 +  \textrm{sin}\frac{k_x}{2} \textrm{sin}\frac{k_y}{2} \textrm{sin}{k_z}(\lambda_{12}  \tau_{1} \sigma_{2}), \nonumber 
\end{eqnarray}
where $\alpha$, $\beta$, $\gamma$ and $\lambda_{ij}$ are hopping strength, and $\delta_0$ represents the chemical potential . As discussed in detail in Ref. \cite{Xiaoting2020a}, the CICE emerge on the mirror plane (gray shaded area in Fig.~\ref{Fig1}a) centered at the high symmetry $k$ point $S=(\pi,\pi,0)$ [$R=(\pi,\pi,\pi)$], under the conditions $\lbrace \alpha \delta_{S (R)} <0 \ \cap \ \alpha \beta > 0 \ \cap \ \alpha \neq \beta \rbrace  $, where $\delta_{S,R} = \delta_0 - (\alpha +\beta \mp \gamma)$. The various terms in Eq. (\ref{H0}) describe the pair of concentric elliptic NLs, the NL anisotropy, and the angle between the NLs (see details in \cite{Xiaoting2020a}). Since the CICE are composed of two NLs, it is anticipated to observe a pair of DSS~\cite{Burkov2011a} intertwined on the (001) surface.

\textbf{Topological crystalline insulator.} We consider the effect of SOC, \textcolor{black}{and the minimal TB model contains 8 spinful bands}. The minimal SOC Hamiltonian in SG $Pbam$ (No. 55), $H_{SOC}(\mathbf{k})$, to gap out the CICE-NL is of the form, 
\begin{eqnarray}\label{HSOC}
&& H_{SOC}(\mathbf{k})   =  \sum^3_{i=1}\zeta_{01i}\textrm{sin}k_{i} \Gamma_{01i} \\
& + & \zeta_{233} \textrm{cos}\frac{k_x}{2} \textrm{cos}\frac{k_y}{2} \textrm{cos}{k_z} \Gamma_{233} 
 + \zeta_{111} \textrm{sin}\frac{k_x}{2} \textrm{cos}\frac{k_y}{2} \textrm{cos}{k_z}  \Gamma_{111} \nonumber \\
& + & \zeta_{112}  \textrm{cos}\frac{k_x}{2} \textrm{sin}\frac{k_y}{2} \textrm{cos}{k_z} \Gamma_{112} 
 + \zeta_{223} \textrm{sin}\frac{k_x}{2} \textrm{sin}\frac{k_y}{2} \textrm{sin}{k_z} \Gamma_{223},\nonumber 
\end{eqnarray}
where $\zeta_{ijk}$ denotes SOC strength, and $\Gamma_{ijk} = \tau_i \sigma_j s_k$ ($i,j,k \in \lbrace 0,1,2,3\rbrace$). $s_{0,1,2,3}$ are identity matrix and Pauli matrices operating in spin space, respectively. The band structure of $H_0$ and $H =H_{0}+H_{SOC}$ are shown in Fig.~\ref{Fig1}d by the red and blue curves, respectively. In the presence of SOC, the CICE-NL TSM evolves into an insulating phase. The parameters are tuned to allow the system to host a single CICE centered at the $S$ point in the absence of SOC and to have no additional band inversions at \textcolor{black}{other k points} including SOC. In addition to $\lbrace \alpha \delta_{S} <0 \cap \alpha \beta > 0 \cap \alpha \neq \beta \rbrace $, either condition $\lbrace \delta_S < 0 \cap \delta_R  > 0 \cap \delta_{\Gamma, Z} > \sqrt{\lambda_{+}^2 + \zeta_{233}^2 } + \sqrt{\lambda_{-}^2 + \zeta_{233}^2 } \rbrace$ or $\lbrace \delta_S > 0 \cap \delta_R  < 0 \cap  \delta_{\Gamma, Z} < -\sqrt{\lambda_{+}^2 + \zeta_{233}^2 } - \sqrt{\lambda_{-}^2 + \zeta_{233}^2 } \rbrace$, should be satisfied, where $\lambda_{\pm} = \lambda_{10} \pm \lambda_{13}$ and $\delta_{\Gamma,Z} = \delta_0 + (\alpha +\beta \pm \gamma)$. \textcolor{black}{If additional band inversions emerge beyond the one which gives rise to the CICE, the band topology may be changed in the presence of SOC, and  the semimetal may evolve to distinct insulating phases, although the CICE nodal lines may still exist in the absence of SOC. }

\begin{figure}
\begin{centering}
\includegraphics[width=\linewidth]{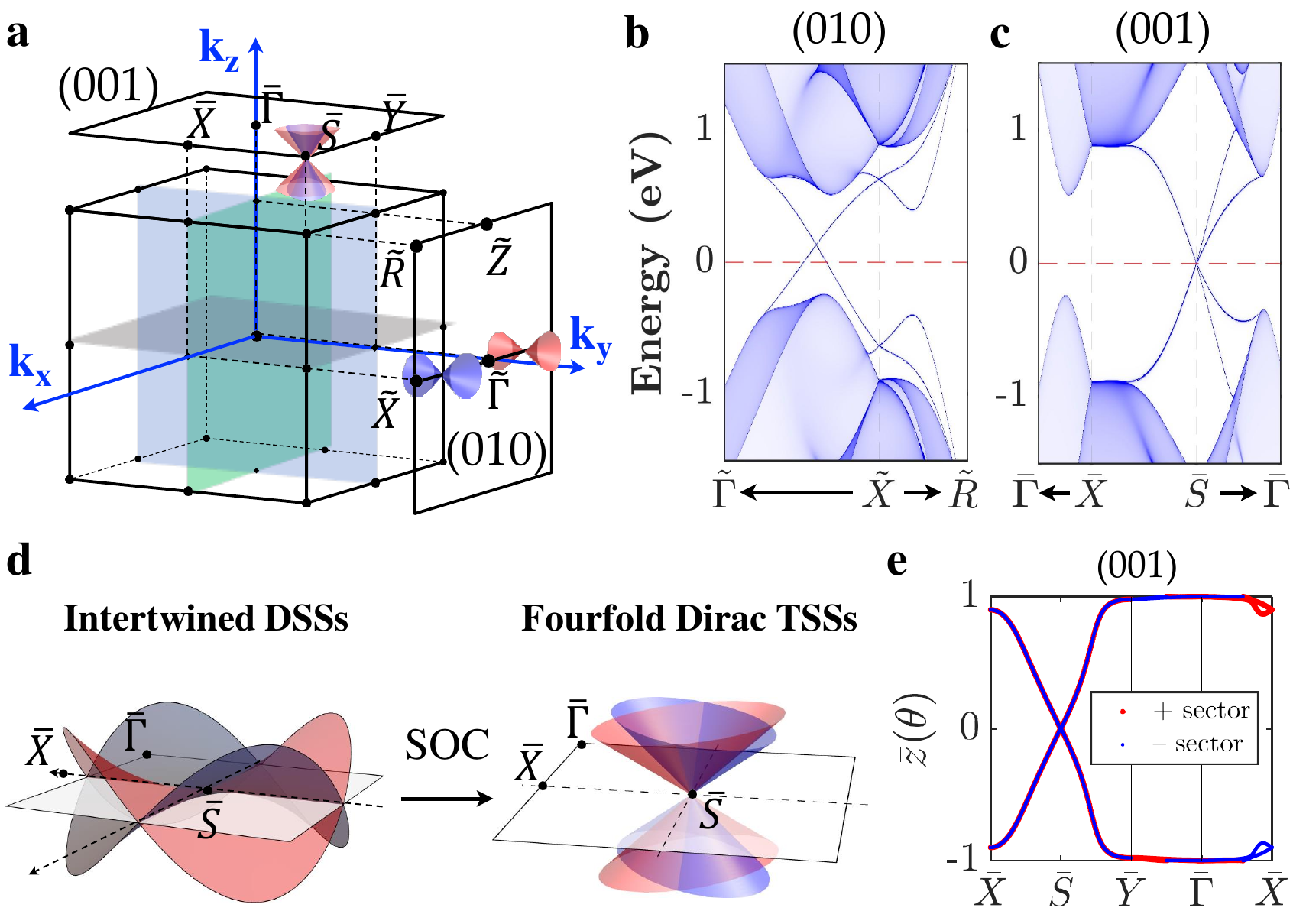}
\par\end{centering}
\centering{}
\caption{ 
{\bf Topological surface states and the schematic transition from DSSs of the NLSM to TSSs of the TCI.} {\bf a,} Brilliouin zones for the bulk, the projected $(001)$ and $(010)$ surfaces, on which the predicted positions of the fourfold Dirac fermion and Dirac cones are shown schematically. Surface band spectrum of the {\bf b,} $(010)$ and {\bf c,} $(001)$ surfaces along the high-symmetry k paths, where the fourfold Dirac fermion emerges at $\bar{S}$ on the (001) surface and a Dirac cone lies along the $\tilde{\Gamma}$ - $\tilde{X}$ direction of the (010) surface, respectively.	
{\bf d,} Schematic of the SOC-driven transition of the (001) surface states from the intertwined DSSs stemming from the CICE NLSM ~\cite{Xiaoting2020a} (left panel) to the fourfold Dirac fermion at $\bar{S}$ of the TCI (right panel).   {\bf e,} The $z$-directed Wilson loop along the high symmetry directions for occupied states hosting positive ($+$ sector, red) and negative ($-$ sector, blue) surface glide eigenvalues, indicating a bulk topology of $(\chi_x, \chi_y)=(2, 2)$.}
\label{Fig2}
\end{figure}

In order to determine its band topology, we implement the symmetry-indicator theory~\cite{Po2017,Khalaf2017a,Song2018b}. Crystals in the SG $Pbam$ (No. 55) are characterized by four symmetry indicators (SIs)~\cite{Po2017,Khalaf2017a,Song2018b}, three $\Z_2$ weak TI indices and one $\Z_4$ index. The $\Z_4$ index is defined as, $\Z_4  \equiv  \dfrac{1}{4}\sum_{K\in TRIMs} (n_{K}^{+} - n_{K}^{-})  \ \textrm{mod} \ 4$, where $n_{K}^{+}$ ($n_{K}^{-}$) is the number of occupied bands with parity $+$ ($-$) at the TRIM points $K$. Due to the nonsymmorphic symmetries, bands are four-fold degenerate at all TRIM points except $\Gamma$. Besides, since inversion $\I$ anticommutes with the glide, $\mathcal{G}$, or screw, $\mathcal{S}$, symmetry operations (here $\mathcal{S}$ includes $\Sy = 2_1^{[010]}= \lbrace 2_{010}\vert\frac{1}{2}\frac{1}{2}0 \rbrace$, and $\Sx = 2_1^{[100]}= \lbrace 2_{100}\vert\frac{1}{2}\frac{1}{2}0 \rbrace$), at X, U, Y, and T, the parity of each four-fold degenerate state must be $(+,+,-,-)$, which does not contribute to $\Z_4$. By enumerating the parity of the states at other TRIM points, we obtain $\Z_{2,2,2,4} = (0,0,0,2)$, corresponding to eight possible topological states~\cite{Song2018b}. To further narrow down the possible phases, we have calculated the mirror Chern numbers of $\Mz$ and $C_{m_{0,\pi}^{001}}$ ($m_{0, \pi}^{001}$ denotes the $k_z =0,\;\pi$ mirror planes), following the method implemented in Ref.~\cite{Zhou2018a}. We find that $C_{m_{0,\pi}^{001}} =(2,0)$. The corresponding Dirac surface states on the (010) surface are shown in Fig.~\ref{Fig2}b, where the relevant $k$ points and the schematic locations of the Dirac cones are illustrated in Fig.~\ref{Fig2}a.
Therefore, given that $C_{m_{0,\pi}^{001}} =(2,0)$, there are two possible topological phases, which are listed  in Table~\ref{TableSI}. 
 The first one is $\mathcal{S}$-protected TCI with  $\nu_{2_1^{[010]}} = \nu_{2_1^{[100]}} =1$,
 while the second one is $\mathcal{G}$-protected TCI with nontrivial $\Z_2$ topological invariants $\nu_{g_a^{(010)}}$ and $\nu_{g_b^{(100)}}$ (where $g_a^{(010)}=\Gy =\lbrace m_{010}\vert\frac{1}{2}\frac{1}{2}0 \rbrace$, and $g_b^{(100)}=\Gx=\lbrace m_{100}\vert\frac{1}{2}\frac{1}{2}0 \rbrace$) .  Note that the nontrivial characteristics of the bands agree with the analysis from the elementary band representations (EBRs)~\cite{Bradlyn2017,Bradlyn2017PRE,Aroyo2017}. The physical EBRs for the 2a Wyckoff position~\cite{Aroyo1,Aroyo2,Aroyo2011} require that the parity of $\Gamma$, Z, S and R has the same sign. However, because of the double band inversion at S guaranteed by the CICE-NL, both valence and conduction bands violate the physical EBRs, suggesting the emergence of nontrivial topology.   


\begin{table}
  \begin{center}
    \caption{The two possible topological states with SIs $\Z_{2,2,2,4} = (0,0,0,2)$ and mirror Chern number $C_{m^{(001)}_{0,\pi}}=(2,0)$ in SG $Pbam$ (No. 55) \cite{Song2018b}. The $\mathbb{Z}$ invariants, $C_{m^{(001)}_{0,\pi}}$, are the mirror Chern numbers for the mirror planes $\mathcal{M}_{(001)}$ with $k_z =0,\;\pi$, respectively. All the listed $\nu$'s are $\mathbb{Z}_2$ classified topological invariants. The set $(\nu_0;\nu_1\nu_2\nu_3)$ are the invariants for 3D $\mathbb{Z}_2$ TIs. $\nu_{g_a^{010}}$ and $\nu_{g_b^{100}}$ represent the invariants for the glide symmetries $\Gy$ and $\Gx$, respectively. $\nu_{\mathcal{I}}$ is the inversion $\mathcal{I}$ protected TCI index, where $\nu_{\mathcal{I}}=1$, features the hinge states in a 3D finite geometry preserving $\mathcal{I}$. $\nu_{2^{001}}$, $\nu_{2_1^{010}}$ and $\nu_{2_1^{100}}$ denote the invariants for the rotational and screw symmetries $2_{[001]}$, $2_1^{[010]}$ and $2_1^{[100]}$, respectively.} 
        \setlength{\tabcolsep}{3.6pt}
      \renewcommand{\arraystretch}{0.8}
    \begin{tabular}{cccccccc}
     \hline
     \hline
      $(\nu_0;\nu_1\nu_2\nu_3)$ & $C_{m^{(001)}_{0,\pi}}$ & $\nu_{g_a^{010}}$ & $\nu_{g_b^{100}}$ & $\nu_{\mathcal{I}}$ & $\nu_{2^{001}}$ & $\nu_{2_1^{010}}$ & $\nu_{2_1^{100}}$ \\
      \hline
      (0;000) & (2,0) & 0 & 0 & 1 & 0 & 1 & 1  \\
      (0;000) & (2,0) & 1 & 1 & 1 & 0 & 0 & 0  \\
      \hline
      \hline
    \end{tabular}
  \label{TableSI}
  \end{center} 
\end{table}

To further determine the topological phase uniquely, we have also investigated the $(001)$ surface bands, because the presence of topological surface states on the (001) surface excludes the scenario of $\mathcal{S}$-protected TCI. The $(001)$ surface bands are shown in Fig.~\ref{Fig2}c, where the (001) surface BZ and the corresponding high symmetry $k$ points are displayed Fig.~\ref{Fig2}a. 
Interestingly, the calculations reveal the emergence of nontrivial surface states around $\bar{S}$ which are composed of two intertwined surface Dirac cones, shown schematically in Fig.~\ref{Fig2}d (right panel). We refer to these topological surface states (TSSs)  as fourfold Dirac fermions, which are of a particular type of wallpaper fermions~\cite{Wieder2018a}.
As Fig.\ref{Fig2}d shows, one can also obtain the fourfold Dirac fermions via SOC-induced splitting of the intertwined DSSs of the CICE TSM~\cite{Xiaoting2020a}, where the degenerate dispersions along $\bar{S}-\bar{X}\;(\bar{Y})$ are guaranteed by $\mathcal{G}$. Consequently, the CICE-NL induced topological phase belongs to the $\mathcal{G}$-protected TCI. In the following, we provide more physical insights on this TCI phase and the fourfold Dirac fermions.   

In general, for time reversal symmetric systems, strong topological insulators (STIs) with a single band inversion at one TRIM point can be regarded as an elementary building block of the nontrivial insulating phase~\cite{Khalaf2017a,Fang2019}. For each STI, the topological surface states of the (001) surface can be described by the Hamiltonian, $h_{\mathbf{k}} = k_x s_2 - k_y s_1$~\cite{Khalaf2017a}. The gapless feature of $h_{\mathbf{k}}$ is protected by the TRS operator, $\T =-is_2 \K$, where $\K$ is the complex conjugation operator. For the current case, since there is a DBI at the $\bar{S}$ point, the induced TCI phase can be viewed as two copies of STIs. Accordingly, for the TSSs of the TCI, the only allowed TR invariant mass term takes the form, $M = m \mu_2 \otimes s_3$, where $\mu_{1,2,3}$ are the Pauli matrices acting on the two copies of $h_{\mathbf{k}}$, and $m$ is constant. If $M$ can be prohibited by any spatial symmetry $\Q$, the anomalous gapless surface states will persist, indicating that the existing topology is protected by $\Q$, which can be $\Gx$ and $\Gy$, as derived below. 

At $\bar{S}$, the eigenvalues of $\Gx$ and $\Gy$ are $\pm 1$. To preserve TRS, the only available representations are $\Gx = \mu_2 \otimes s_1$ and $\Gy = \mu_2 \otimes s_2$, which in turn lead to the rotational symmetry about the $z$-axis $C_{2z} (=\Gx \times \Gy) = -i \mu_0 \otimes s_3$. Obviously, $M$ cannot survive with $\Gx$ and $\Gy$, but is allowed by $C_{2z}$. Consequently, there exist representations for $\Gx$ and $\Gy$ to support the (001) TSSs at $\bar{S}$, which is the fourfold Dirac fermion shown in Fig.~\ref{Fig2}d described by the \kp Hamiltonian 
\begin{eqnarray}
H_{TSS} (\mathbf{q})  &=&  g_0 (q_x s_2 - q_y s_1) + \sum_{i=1,3} g_i\mu_i \otimes (q_x s_1 + q_y s_2 ) \nonumber \\
&+& q_x q_y (a_3 \mu_3+ a_1 \mu_1) + g_{23}  \mu_2 s_3,
\label{TSSs}
\end{eqnarray}
where all $g$'s and $a$'s are real parameters.

{The $z$-oriented Wilson loop~\cite{Yu2011,Vanderbilt2014}, $\bar{z}_n^{\pm}(k_x, k_y) = \langle W_{n0}^{\pm}|\hat{z}|W_{n0}^{\pm}\rangle$, is calculated on the (001) surface, where $|W_{n0}^{\pm}\rangle$ is the $n$-th Wannier orbit with glide $\Gx$ ($\Gy$) eigenvalues $\pm e^{-ik_y}$ ($\pm e^{-ik_x}$) in the home unit cell, $R = 0$ and $\hat{z}$ is the position operator. Fig.~\ref{Fig2}e shows $\bar{z}^{\pm}$ along the high symmetry $k$ directions for occupied states hosting positive ($+$ sector, red) and negative ($-$ sector, blue) surface glide eigenvalues. Employing the analysis for the two $\mathbb{Z}_4$ indexes $(\chi_x, \chi_y)$ derived in Ref.~\cite{Wieder2018a}, the bulk topology of our system is $(\chi_x,\chi_y) = (2, 2)$.}

\textbf{Higher-order topological insulator protected by a glide symmetry.} In addition to the topological surface states belonging to the $(d-1)$ bulk-edge correspondence, \textit{i.e.,} the fourfold Dirac fermion and the $\Mz$-protected surface states, the $\Gx$, $\Gy$ and $\Mz$ symmetries can give rise to higher order $(d-2)$ bulk-edge correspondence.

\begin{figure}
\begin{centering}
\includegraphics[width=\linewidth]{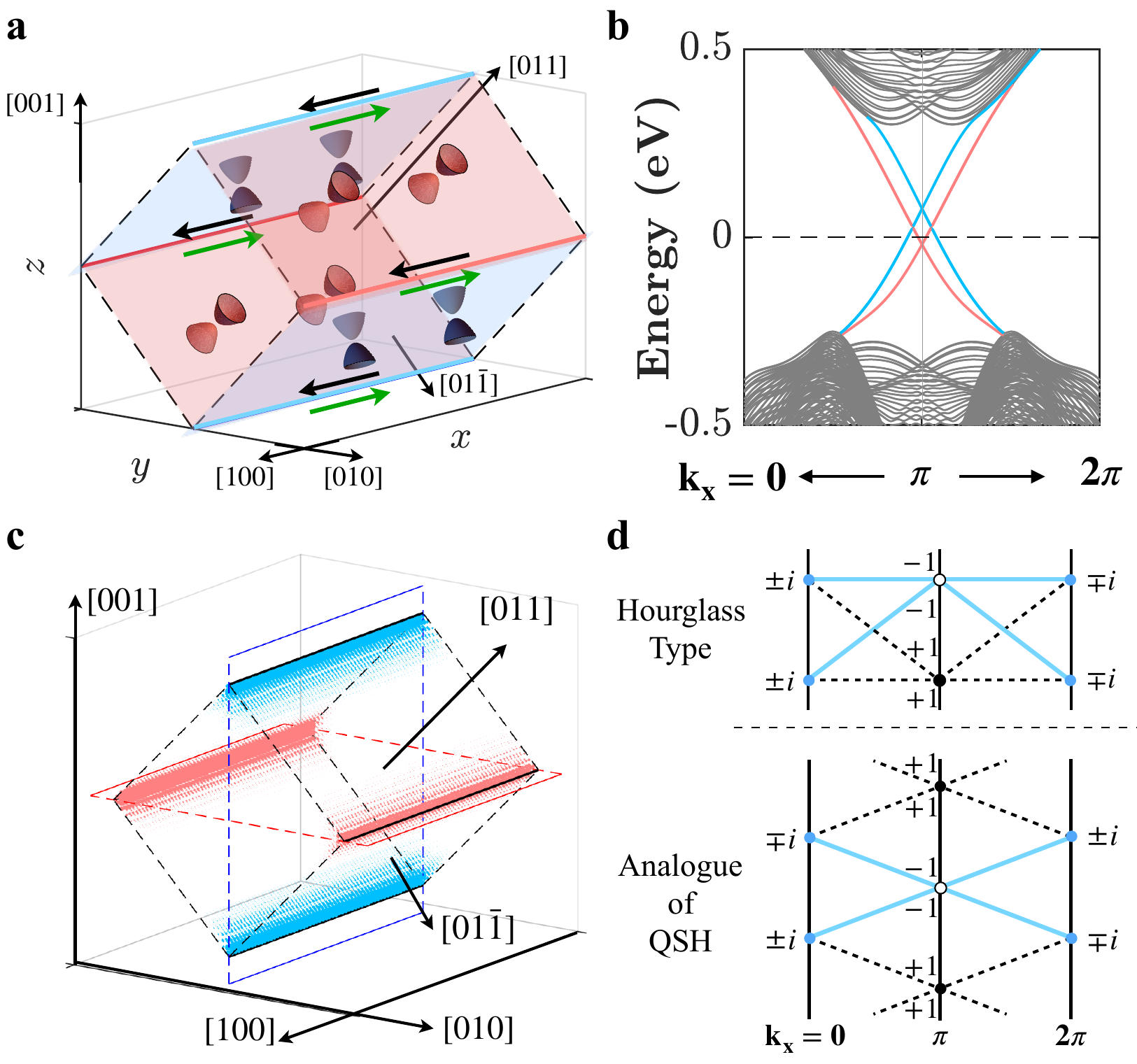}
\par\end{centering}
\centering{}
\caption{
{\bf Bulk-surface-hinge correspondence of the HOTI, and the band connectivity the 1D hinge states protected by a glide symmetry.}  {\bf a,} Geometry of the nanorod, which is periodic along the $[100]$ direction, and finite-size along the $[011]$ and $[01\bar{1}]$ directions. 
Two pairs of hinge modes (denoted in red and blue) emerge along the intersection lines between the $(011)$ and $(0\bar{1}1)$ facets and the $(011)$ and $(01\bar{1})$ facets, respectively.	
The first (second) pair of hinge states, denoted by cyan (red), preserves the glide $\Gy$ (mirror $\Mz$)  symmetry. 
 {\bf b,} Energy dispersion along $k_x$ for the nanorod where the topological hinge modes are denoted with the corresponding cyan and red colors, respectively. 
 {\bf c,} Distribution in real space of the hinge states of the nanorod. For  purpose of clarity, on-site potentials V = 0.2 eV  are added  for the hinges.
 {\bf d,} Schematic of two symmetry-allowed band connectivities of the hinge states (cyan) protected by glide symmetry $\Gy$.}
\label{Fig3}
\end{figure}

We have considered the nanorod geometry, shown in Fig.~\ref{Fig3}a, with open boundary conditions along the $[011]$ and $[01\bar{1}]$ directions, respectively, and periodic boundary conditions along $[100]$. 
 We find that the nanorod can support two pairs of hinges modes along the intersection lines between the $(011)$ and $(0\bar{1}1)$ surfaces and the $(011)$ and $(01\bar{1})$ surfaces, respectively. None of the above surfaces hosts gapless surface states, since there is no TCI phase supporting them. 
As discussed in Ref.~\cite{Schindler2017a}, the hinge modes formed by the intersection of the $(011)$ and $(0\bar{1}1)$ facets are generated via bending the $(001)$ surface along the $[001]$ direction, which, however, preserves the $\Gy$ symmetry for the hinges and the entire crystal.
The original pair of Dirac cones forming the fourfold Dirac fermion on the $(001)$ surface become gapped with opposite mass terms and reside on the $(011)$ and $(0\bar{1}1)$ surfaces, respectively (red and black massive Dirac cones in Fig.~\ref{Fig3}a). {Hence, the surface insulating phases that reside on the two facets differ by an odd $\mathbb{Z}_2$ index, leading to the emergence of an odd number of helical hinge modes on the domain wall between the two facets.} Similarly, the hinge modes along the intersection of the $(011)$ and $(01\bar{1})$ facets are generated via bending the $(010)$ surface along the $[001]$ direction, which preserve the $\Mz$ symmetry.

Fig.~\ref{Fig3}b shows the band structure of the nanorod along the $k_x$ symmetry direction where the two pair of hinge modes are denoted in red and cyan, respectively. The distribution in real space of the two types of hinge modes along the $[100]$ direction are displayed in Fig.~\ref{Fig3}c with the corresponding colors. 
For the purpose of clarity, on-site potentials $V=0.2$ eV are added on the hinges formed by the $\lbrace (011),(0\bar{1}1)\rbrace$ and $\lbrace (0\bar{1}\bar{1}),(01\bar{1})\rbrace$ facets to better differentiate the two hinge states in energy. 
{The constraint imposed by the glide symmetry $\Gy$ allows two possible topologies of hinge bands~\cite{Wang2016b}, the hourglass connectivity and the analogue of the quantum spin Hall (QSH) effect (see Fig.~\ref{Fig3}d). From the the hinge modes in cyan displayed in Fig.~\ref{Fig3}c, we find that the hinge band connectivity is the analogue of the quantum spin Hall effect  (bottom panel of Fig.~\ref{Fig3}d).} 

\section{Discussion}
In Ref.~\cite{Xiaoting2020a}, we demonstrated that 9 ($Pbam$ (No. 55), $Pccn$ (No. 56), $Pnnm$ (No. 58), $Pnma$ (No.62),$P4/mbm$ (No. 127), $P4/mcn$ (No.128), $P4_2/mbc$ (No.135), $P4_2/mnm$ (No.136), $P4_2/ncm$ (No.138)) out of 230 SGs can host CICEs centered at certain TRIM points in the absence of SOC. However, only two wallpaper groups, namely $pgg$ ((001) surface of $Pbam$ (No. 55)) and $p4g$ ((001) surface of $P4/mbm$ (No. 127) and $P4_2/mbc$ (No. 135)), containing double-glide lines, can support fourfold Dirac fermions when SOC is introduced. Consequently, only 3 ($Pbam$ (No. 55), $P4/mbm$ (No. 127) and $P4_2/mbc$ (No. 135)) of the 9 SGs can support the fourfold Dirac fermions on their (001) surfaces. In SG $Pnma$ (No.55), the CICE resides on the (010) rather than the (001) plane, and the (010) surface ($pg$) contains a single glide line. Therefore, the (010) TSSs of the corresponding TCI are hourglass fermions instead. For the rest 5 SGs, due to the lack of glide symmetry preserved on their (001) surfaces, the nonsymmorphic fourfold Dirac or hourglass fermions would not be expected.

In summary, inclusion of SOC in the model Hamiltonian describing our recently proposed a family of butterfly-like CICE NLs in SG $Pbam$ (No. 55)~\cite{Xiaoting2020a} unveils intrinsic connection of the CICE NLSM and the TCI  protected by two glide symmetries. The SOC drives the TSM to a $\Z_4=2$ TCI with higher order topology, supporting in turn a fourfold Dirac fermion on the (001) surface protected by two coexisting glide symmetries  of WG $pgg$. 
As a candidate material of this type of TCI, 
Sr$_2$Pb$_3$ (SG $P4/mbm$ No.127) has been studied in~\cite{Wieder2018a}. 
Moreover, its higher order topology is corroborated for the first time by the emergence of 1D hinge states protected by glide symmetry. This intriguing TCI phase provides a platform for exploring exotic physics, such as the electron transport and thermoelectric effect on the surfaces/hinges. Our proposed glide-protected HOTI may have important implications on the emergence of Majorana zero modes via proximity of the HOTI to a superconductor~\cite{Hsu2018}, which exhibit distinct features compared to those in TI/SC heterostructures.  {Finally, our EBRs analysis of the tight binding model in Supplemental Materials~\cite{JuanSI} (Section. B) demonstrates a rich phase diagram featuring TCI,  strong topological insulator (TI)  and obstructed atomic insulator (OAI) phases.}

\textcolor{black} 
{As described in detail in the Supplemental Material, we have carried out systematic {\it ab initio} electronic structure calculations to identify material candidates 	which exhibit (i) a {\it single} CICE in the absence of SOC and (ii) fourfold-Dirac TCI/HOTI phase in the presence of SOC. Unfortunately, this task has proved immensely challenging and the calculations failed in finding such an ideal material. This is due to the fact that the band structures in real materials are  more complex involving multiple orbitals and bands. (See details in  in Supplemental Materials~\cite{JuanSI} (Section. A)) Nevertheless, these calculations raise the intriguing question of realizing these quantum states by designing new materials according to our simple model, so that the topological nature can be captured and not buried in the complex band structure. 
	As stated above, we have constructed a minimal (8-band in the presence of SOC) tight-binding model with a bipartite lattice in space group Pbam (No. 55), based on the Wyckoff position 2a with site-symmetry group $2/m$, and with two orbital per sublattice. 
Even though we employed the $(p_z, d_{xy})$-derived orbitals, this choice is not unique and one can equally well use the $(p_z, s)$-derived orbitals. 
The conditions for the emergence of CICE nodal lines in the absence of SOC are given by Eqs. (4) and (5) of Ref.~\cite{Xiaoting2020a} with the additional proviso that no other band inversions occur. The conditions to avoid additional band inversions in the presence of SOC are given in the paragraph below Eq.~\ref{HSOC}) of this work. The artificial system can be designed as layered structure, where these conditions can be satisfied by tuning the interlayer hopping or SOC parameters, such as $\gamma$, $\lambda_{10,13}$ and $\zeta_{233}$.}

\section{Acknowledgements}
The work at CSUN was supported by NSF-Partnership in Research and Education in Materials (PREM) Grant No. DMR-1828019. H.L. acknowledges the support by the Ministry of Science and Technology (MOST) in Taiwan under grant number MOST 109-2112-M-001-014-MY3. The work of J.L.M. has been supported by Spanish Science Ministry grant PGC2018-094626-B-C21 (MCIU/AEI/FEDER, EU) and Basque Government grant IT979-16. M.G.V. thanks support from DFG INCIEN2019-000356 from Gipuzkoako Foru Aldundia. M.G.V. and M.I. acknowledges the Spanish Ministerio de Ciencia e Innovacion (grant number PID2019-109905GB-C21).

\section{Competing Interests}
The authors declare that they have no competing financial interests.

\section{Author Contributions}
X.Z. designed research. X.Z., C.-H.H, M.I., C.-Y.H., J.M.L., M.G.V., H.L. and N.K. performed research. X.Z., C.-H.H, M.I., C.-Y.H., J.M.L. analyzed data. X.Z. and C.-H.H. drafted the main text, whereas J.M.L, M.I., and X.Z. drafted the supplementary information. All authors contributed to the editing of the manuscript.



\newpage

\newpage

\newpage

\clearpage
\widetext
\setcounter{equation}{0}
\setcounter{figure}{0}
\setcounter{table}{0}
\renewcommand{\theequation}{S\arabic{equation}}
\renewcommand{\thefigure}{S\arabic{figure}}
\renewcommand{\thetable}{S\arabic{table}}
\renewcommand{\bibnumfmt}[1]{[S#1]}
\renewcommand{\citenumfont}[1]{S#1}
\newcommand{\bk}{\boldsymbol\kappa}

\newcommand{\SI}{Supplementary Material}
\newcommand{\beginsupplement}{%
  \setcounter{equation}{0}
  \renewcommand{\theequation}{S\arabic{equation}}%
  \setcounter{table}{0}
  \renewcommand{\thetable}{S\arabic{table}}%
  \setcounter{figure}{0}
  \renewcommand{\thefigure}{S\arabic{figure}}%
  \setcounter{section}{0}
  \renewcommand{\thesection}{S\Roman{section}}%
  \setcounter{subsection}{0}
  \renewcommand{\thesubsection}{S\Roman{section}.\Alph{subsection}}%
}

\def\y{\'{\i}}
\def\to{\rightarrow}
\def\l{\langle}
\def\r{\rangle}
\def\p{\partial}
\def\q{{?`}}
\def\ni{\noindent}
\def\e{\varepsilon}
\def\d{\textrm{d}}
\def\non{\nonumber }
\def\D{\Delta}
\def\om{\omega}
\def\ra{\rangle}
\def\la{\langle}
\def\s{\sigma}
\def\med{\frac{1}{2}}

\newcommand{\beqa}{\begin{eqnarray}} 
\newcommand{\eeqa}{\end{eqnarray}}

\section{Supplemental Material}

\subsection{EBR Analysis of the Tight-Binding Model for $SG55$}

The hamiltonian given by Eqs. (1-2) in the Main Text can be obtained as a tight-binding model based on the Wyckoff position $2a$ with site-symmetry group $2/m$. Two orbitals  $(p_z,d_{xy})$ are placed at  the origin of the cell $\mathbf{r}_A=(0,0,0)$ and at $\mathbf{r}_B=(1/2,1/2,0)$. The Hamiltonian $H(\mathbf{k})=H_0(\mathbf{k})+H_{SOC}(\mathbf{k})$ is invariant under the generators of $SG55$ given by
\beq
\{ C_{2z}|0,0,0\}=\tau_0\otimes\sigma_0 \otimes s_3\;,\; \{ C_{2x}|\med\med 0\}=e^{{\frac {i}{2}(k_y-k_x)}}\tau_1\otimes\sigma_0\otimes s_1\;,\;
\{ I|000\}=\tau_0\otimes\sigma_3 \otimes s_0 ,
\eeq
where $\tau_0$, $\sigma_0$ and $s_0$ are $2\times 2$ identity matrices. The matrices $\tau_i$ act on  the  $p_z$ and $d_{xy}$ orbitals, $\sigma_i$ on the  two sublattices and $s_i$ on electron spin. Note that the invariance of the hamiltonian under a space-group operation $g=\{ R| \mathbf{v}\}$  means that

\beq
g H_0(\mathbf{k})g^\dagger=H_0(R\mathbf{k}).
\eeq
Similarly, the invariance of the hamiltonian under the anti-unitary time-reversal operation $\mathcal{T}=T\mathcal{K}$  can be expressed as
\beq
T H_0^*(\mathbf{k})T^\dagger=H_0(-\mathbf{k}),
\eeq
where   $T=-i\tau_0\otimes\sigma_0 \otimes s_2$ and $\mathcal{K}$ is ordinary complex conjugation.

\vskip1cm
\ni{\bf FROM ATOMIC ORBITALS TO EBRS}
\vskip.5cm
Roughly speaking, a band representation~\cite{zak1,zak2} can be understood as the collection of electronic bands  generated by placing a set orbitals at different positions in the unit cell, in such a way that the set is closed under all the symmetries of the crystal.  A band representation obtained by placing orbitals at a maximal Wickoff position $w$ that transform according to an irreducible representation $\tau$ of the site symmetry group $G_w$ is an \textit{elementary} band representation (EBR) and 
is denoted by $\tau\uparrow G|_w$. Any band representation that is not elementary  can be written as a sum of EBRs~\cite{ebr3}.

 As our system respects time reversal symmetry (TRS), we are interested only in real and physically irreducible representations~\cite{brad}. Following the Bilbao Crystallographic Server~\cite{mois1,mois2} (BCS) conventions,  a physically irreducible representation is denoted as the sum of two  irreducible representations, which can be the same or different depending on their reality type~\cite{brad}, with the $+$ sign between the components of the pair  omitted. For instance, $\Gamma_1^+$ in Eq.~(\ref{AgAu}) below is a real irreducible representation whereas $R_1^+$ and $R_2^+$ are complex conjugate irreducible representations that together constitute the real, physically irreducible representation $R_1^+R_2^+$.

In order to find the EBR contents of the model it is enough to know the transformation properties of the atomic orbitals $s$ and $p_z$. The site-symmery group $2/m$ has four single-valued real irreducible representations (irreps), $A_g,A_u,B_g,B_u$, and two double-valued physically irreducible irreps, $
^1\!\bar E_g\, ^2\!\bar E_g$ and $^1\!\bar E_u\, ^2\!\bar E_u$. The spinless  orbitals $d_{xy}$ and $p_z$ belong the single-valued irreps $A_g$ and $A_u$ respectively. As a consequence, in the absence of SOC the spectrum of the TB hamiltonian (5)  is described by the sum of two EBRs induced from the $2a$ Wyckoff position, namely
\beq
A_g\!\uparrow\! G|_{2a}+A_u\!\uparrow\! G|_{2a}.
\eeq

Then the application \textit{BANDREP} at the BCS gives all the irreps at the different points in the Brillouin Zone (BZ). For the high symmetry TRIM points the result is
\beqa\label{AgAu}
A_g\!\uparrow\! G|_{2a}&=&\{\Gamma_1^+(1),\Gamma_2^+(1);R_1^+R_2^+(2);S_1^+S_2^+(2);T_2(2);U_1(2),X_1(2),Y_2(2),Z_1^+(1),Z_2^+(1)\}\non \\
A_u\!\uparrow\! G|_{2a}&=&\{\Gamma_1^-(1),\Gamma_1^-(1);R_1^-R_2^-(2);S_1^+S_2^+(2);T_1(2);U_2(2),X_2(2),Y_1(2),Z_1^-(1),Z_2^-(1)\},
\eeqa
where the numbers in parentheses give the dimensions of the irreps, which coincide with the degeneracies of the corresponding bands. This means, for instance, that the hamiltonian  for spinless electrons in Eq.~(5) of the paper necessarily has four non-degenerate bands at the  $\Gamma$-point labelled by the 1-dim irreps $\{\Gamma_1^+(1),\Gamma_2^+(1),\Gamma_1^-(1),\Gamma_2^- (1)\}$, two doubly-degenerate bands 
at the $R$-point labelled by the 2-dim physically irreps $R_1^+R_2^+(2)$ and $R_1^-R_2^-(2)$, etc. Note that in order to obtain this information one does not have to diagonalize the hamiltonian $H_0(\mathbf{k})$, as  the result depends only on the orbital contents of the model. Note also that the degeneracies of the bands at the high symmetry points are independent of the parameters of the model, barring accidental degeneracies if some parameters are set to zero or otherwise fine-tuned. 

Electron spin and SOC can be easily incorporated by noting that spin transforms according to the physically irrep $^1\!\bar E_g\, ^2\!\bar E_g$ of the site-symmetry group $2/m$~\cite{brad}. Then, the double-valued irreps for $s$ and $p$ orbitals are obtained by taking the products of the respective single-valued irreps with the spin representation
\beq 
A_g\times  ^1\!\!\bar E_g\, ^2\!\bar E_g =\, ^1\!\bar E_g\, ^2\!\bar E_g\;\;,\;\; A_u\times ^1\!\!\bar E_g\, ^2\!\bar E_g =\, ^1\!\bar E_u\, ^2\!\bar E_u,
\eeq
and this implies that the spectrum of the total hamiltonian $H(\mathbf{k})=H_0(\mathbf{k})+H_{SOC}(\mathbf{k})$ is described by the sum of two EBRs
\beq\label{sumebrs}
^1\!\bar E_g\, ^2\!\bar E_g\uparrow\! G|_{2a} + ^1\!\!\bar E_u\, ^2\!\bar E_u\uparrow\! G|_{2a}.
\eeq
Then \textit{BANDREP} immediately gives the irrep contents and the structure of the spectrum for the total hamiltonian $H(\mathbf{k})$. For the high symmetry TRIM points the result is 
\beqa\label{2ebrs}
^1\!\bar E_g\, ^2\!\bar E_g\uparrow\! G|_{2a}&=&\{2\bar \Gamma_5(2);\bar R_5\bar R_5(4);\bar S_5\bar S_5(4);\bar T_3\bar T_4(4);\bar U_3\bar U_4(4);\bar X_3\bar X_4(4);\bar Y_3\bar Y_4(4);2\bar Z_5(2)\}\non\\
^1\!\bar E_u\, ^2\!\bar E_u\uparrow\! G|_{2a}&=&\{2\bar \Gamma_6(2);\bar R_6\bar R_6(4);\bar S_6\bar S_6(4);\bar T_3\bar T_4(4);\bar U_3\bar U_4(4);\bar X_3\bar X_4(4);\bar Y_3\bar Y_4(4);2\bar Z_6(2)\}.
\eeqa

\vskip.5cm
\ni{\bf FROM EBRS TO  SYMMETRY-BASED INDICATORS}
\vskip.5cm
The irreps at all the high symmetry points in the Brillouin zone (BZ) are tabulated at the BCS. For crystals with SOC and time reversal symmetry (TRS), we are interested in double-valued (because of the SOC), physically irreducible  (because of TRS) representations. For SG55 there are $12$ double-valued, physically irreducible representations
\beq\label{pirreps}
\{\bar \Gamma_5(2),\bar \Gamma_6(2);\bar R_5\bar R_5(4),\bar R_6\bar R_6(4);\bar S_5\bar S_5(4),\bar S_6\bar S_6(4);\bar T_3\bar T_4(4);\bar U_3\bar U_4(4);\bar X_3\bar X_4(4);\bar Y_3\bar Y_4(4);\bar Z_5(2),\bar Z_6(2)\}.
\eeq 
Then any band representation can be characterized by a vector that tells how many times each physically irreducible representation appears at every high symmetry point. For instance, for the two EBRs in Eq.~(\ref{2ebrs}) the vectors are
\beqa\label{vectors}
^1\!\bar E_g\, ^2\!\bar E_g\uparrow\! G|_{2a}&=&(2, 0, 1, 0, 1, 0, 1, 1, 1, 1, 2, 0)\non\\
^1\!\bar E_u\, ^2\!\bar E_u\uparrow\! G|_{2a}&=&(0, 2, 0, 1, 0, 1, 1, 1, 1, 1, 0, 2),
\eeqa
and for the bands of the total hamiltonian $H(\mathbf{k})=H_0(\mathbf{k})+H_{SOC}(\mathbf{k})$
\beq\label{comp8}
^1\!\bar E_g\, ^2\!\bar E_g\uparrow\! G|_{2a} + ^1\!\!\bar E_u\, ^2\!\bar E_u\uparrow\! G|_{2a}=(2, 2, 1, 1, 1, 1, 2, 2, 2, 2, 2, 2).
\eeq

The topology of an \textit{isolated} subset of bands can be characterized by a set of symmetry-based indicators~\cite{indi}. In order to determine the indicators, one tries to write the vector giving the irreps for the subset of bands as a linear combination of all the vectors for the EBRs of the space group. There are several possibilities:
\begin{enumerate}
\item If the vector for the subset of bands can be written as a linear combination with \textit{positive integer} coefficients, then all symmetry-based indicators (SI) for the subset vanish. The subset may still be topologically non-trivial, but all the eigenvalue-based topological invariants will be zero. In order to ascertain that the subset is non-trivial, one should use other tools, such as Wilson loops. 

\item If the vector for the subset of bands can be written as a linear combination with \textit{integer} coefficients, but some of the coefficients are necessarily negative, then the subset exhibits non-trivial \textit{fragile} topology.

\item If the vector for the subset of bands can not be written as a linear combination with \textit{integer} coefficients, then there are SI of the form
$Z_{n_1}\times Z_{n_2}\times\ldots$, where the $Z_{n}$-factors reflect the existence of \textit{fractional coefficients} in the linear combination of EBRs. These subsets of bands are usually listed as NLC, meaning non-linear combination of EBRs with integer coefficients, and are characterized by the existence of different topological invariants, such as $Z_2$, $Z_4$, etc.

\item If the vector for the subset of bands can not be written as a linear combination at all, then the subset is not really isolated, i.e., the gap between the subset and the other bands in the crystal closes at some point in the Brillouin zone.

\end{enumerate}

The group of symmetry-based indicators for a given space group can be determined by diagonalization of  the matrix $EBR$. This matrix is constructed by taking as columns the vectors of all the EBRs for the group.  A look at \textit{BANDREP} shows that, with TRS,  there are eight double-valued EBRs for $SG55$. The first two have been given in Eq.~(\ref{vectors}). Taking all $8$ vectors as columns gives the $EBR$ matrix for $SG55$
\beq
EBR=\left(
\begin{array}{cccccccc}
 2 & 0 & 2 & 0 & 2 & 0 & 2 & 0 \\
 0 & 2 & 0 & 2 & 0 & 2 & 0 & 2 \\
 1 & 0 & 0 & 1 & 0 & 1 & 1 & 0 \\
 0 & 1 & 1 & 0 & 1 & 0 & 0 & 1 \\
 1 & 0 & 1 & 0 & 0 & 1 & 0 & 1 \\
 0 & 1 & 0 & 1 & 1 & 0 & 1 & 0 \\
 1 & 1 & 1 & 1 & 1 & 1 & 1 & 1 \\
 1 & 1 & 1 & 1 & 1 & 1 & 1 & 1 \\
 1 & 1 & 1 & 1 & 1 & 1 & 1 & 1 \\
 1 & 1 & 1 & 1 & 1 & 1 & 1 & 1 \\
 2 & 0 & 0 & 2 & 2 & 0 & 0 & 2 \\
 0 & 2 & 2 & 0 & 0 & 2 & 2 & 0 \\
\end{array}
\right).
\eeq
Note that $EBR$ is a $12\times 8$ non-square matrix, where $8$ is the number of EBRs and $12$ the number of irreps. Then, for any isolated subset of bands characterized by a $12$-component vector $B=(B_1,\ldots,B_{12})$ giving the numbers of irreps, we will have to solve the linear set of equations
\beq
B=EBR\cdot X,
\eeq
where the solution $X=(X_1,\ldots,X_{8})$ gives the coefficients of the linear combination of the $8$ EBRs. The presence of non-integer the coefficients in $\{X_1,\ldots,X_{8}\}$
implies the existence of symmetry-based indicators. 

The possible existence of non-integer solutions and hence the  existence of SI and non-trivial topological indicators may be predicted by diagonalizing the $EBR$ matrix. As $EBR$ is a non-square matrix with integer coefficients, one has to use a special diagonalization procedure known as \textit{Smith decomposition}:
\beq
D=L\cdot EBR\cdot R,
\eeq
where $L$ and $R$ are unitary integer square matrices of dimensions $12\times 12$ and $8\times 8$ respectively, and $D$ is the following $12\times 8$ diagonal matrix
\beq
D=\left(
\begin{array}{cccccccc}
 1 & 0 & 0 & 0 & 0 & 0 & 0 & 0 \\
 0 & 1 & 0 & 0 & 0 & 0 & 0 & 0 \\
 0 & 0 & 1 & 0 & 0 & 0 & 0 & 0 \\
 0 & 0 & 0 & 2 & 0 & 0 & 0 & 0 \\
 0 & 0 & 0 & 0 & 4 & 0 & 0 & 0 \\
 0 & 0 & 0 & 0 & 0 & 0 & 0 & 0 \\
 0 & 0 & 0 & 0 & 0 & 0 & 0 & 0 \\
 0 & 0 & 0 & 0 & 0 & 0 & 0 & 0 \\
 0 & 0 & 0 & 0 & 0 & 0 & 0 & 0 \\
 0 & 0 & 0 & 0 & 0 & 0 & 0 & 0 \\
 0 & 0 & 0 & 0 & 0 & 0 & 0 & 0 \\
 0 & 0 & 0 & 0 & 0 & 0 & 0 & 0 \\
\end{array}
\right).
\eeq
The number of non-zero diagonal elements   is the \textit{rank} of the group, equal to five in the case of $SG55$ with SOC and TRS. One can show that all the eigenvalues must be positive integers. If some of the eigenvalues$\{n_1, n_2,\cdots\}$ are greater than one, then the group of SI is $Z_{n_1}\times Z_{n_2}\times\ldots$. In this case the SI group is $Z_2\times Z_4$, in agreement with Table~III of Ref.~\onlinecite{indi}. Note that, upon inversion, the existence of greater than one eigenvalues may give rise to non-integer values for some of the coefficients of the solution $X$, showing that we are in case 3 of the previous discussion.

\vskip.5cm
\ni{\bf COMPUTING THE SYMMETRY INDICATORS FOR THE TB MODEL}
\vskip.5cm
According to Table~IV of Ref.~\onlinecite{indi}, SG55 has no SI in the absence of SOC and, as a consequence, can not have 
eigenvalue-based topological indices. For that reason we will consider the complete hamiltonian $H(\mathbf{k})=H_0(\mathbf{k})+H_{SOC}(\mathbf{k})$.
On the other hand,  Eq.~(\ref{sumebrs}) shows that the total eight-band spectrum can be written as the sum of two EBRs. Thus, in order to find non-trivial topology, 
 we have to consider the possibility of splitting the eight bands into disconnected subsets, in such a way that the subsets have non-trivial SI. Put differently, if the two EBRs in  Eq.~(\ref{2ebrs}) represent the valence and conduction bands, both will be topologically trivial. We may get non-trivial topology only through inversions of energy levels between the conduction and valence bands. 
 
 A look at Eq.~(\ref{2ebrs}) shows that inversions can  take place only at the $\Gamma$, $R$, $S$ and $Z$ points. There are obviously three ways to split the irreps among the conduction and valence bands at the $\Gamma$ and $Z$ points, two at  $R$ and $S$ and just one at the remaining high symmetry points. This gives a total of $3\times 3\times 2\times 2 =36$ patterns. The possible irrep contents for the different subsets of four bands,  together with the computed topological invariants,  are shown in Table~\ref{t1} using the vector notation explained below Eq.~(\ref{pirreps}). The conduction band for phase considered in the Main Text with a single double band inversion  at the $S$-point corresponds to case 4 in the Table~\ref{t1}.
 
 Note that only half the possible subsets  have been listed. The remaining subsets are \text{complementary} to the ones given there, and so are their SI.
 Considering for instance the third entry in the table, given by \hbox{$B=(2, 0, 1, 0, 1, 0, 1, 1, 1, 1, 1,1)$}, the complementary band is obtained by subtracting it from  the complete set of bands of the model, as given in Eq.~(\ref{comp8}) 
\beq
(2, 2, 1, 1, 1, 1, 2, 2, 2, 2, 2, 2)-(2, 0, 1, 0, 1, 0, 1, 1, 1, 1, 1,1)=(0, 2, 0, 1, 0, 1, 1, 1, 1, 1, 1,1)
\eeq
and its SI are 
\beq
Z_2=Z_{2w,1}=1\;,\; Z_{2w,2}=Z_{2w,3}=0\;,\; Z_4=1.
\eeq
Adding the topological invariants of the two complementary bands and remembering that $Z_n$ indicators are defined $mod$ n, gives for the total 8-band system
\beq
Z_2=Z_{2w,1}=Z_{2w,2}=Z_{2w,3}= Z_4=0
\eeq
as expected. Note that the conduction and valence bands are always complementary.

 \begin{table}[t]
\begin{tabular}{|c|| c| c|c |c|c| c | c |}
\hline
No &B & $Z_2$ & $Z_{2w,1}$ & $Z_{2w,2}$ & $Z_{2w,3}$ & $\;Z_{4}\;$& EBR\\ 
 \hline
 \hline
1&$(2, 0, 1, 0, 1, 0, 1, 1, 1, 1, 2, 0)$ & $0$ & $0$ & $0$ & $0$ & $0$ & $^1\!\bar E_g \,^2\!\bar E_g\uparrow G|_{2a}$\\
2&$(2, 0, 1, 0, 1, 0, 1, 1, 1, 1, 0, 2)$ & $0$ & $0$ & $0$ & $0$ & $2$&-\\
3&$(2, 0, 1, 0, 1, 0, 1, 1, 1, 1, 1,1)$ & $1$ & $0$ & $0$ & $1$ & $3$&-\\
 \hline
4& $(2, 0, 1, 0, 0,1, 1, 1, 1, 1, 2, 0)$ & $0$ & $0$ & $0$ & $0$ & $2$&-\\
5&$(2, 0, 1, 0, 0,1, 1, 1, 1, 1, 0, 2)$ & $0$ & $0$ & $0$ & $0$ & $0$& $^1\!\bar E_g \,^2\!\bar E_g\uparrow G|_{2d}$\\
6&$(2, 0, 1, 0, 0,1, 1, 1, 1, 1, 1,1)$ & $1$ & $0$ & $0$ & $1$ & $1$&-\\
 \hline
 7&$(2, 0, 0,1, 1,0, 1, 1, 1, 1, 2, 0)$ & $0$ & $0$ & $0$ & $0$ & $2$&-\\
8&$(2, 0, 0,1, 1,0, 1, 1, 1, 1, 0, 2)$ & $0$ & $0$ & $0$ & $0$ & $0$& $^1\!\bar E_g \,^2\!\bar E_g\uparrow G|_{2b}$\\
9&$(2, 0, 0,1, 1,0, 1, 1, 1, 1, 1,1)$ & $1$ & $0$ & $0$ & $1$ & $1$&-\\
 \hline
 10&$(2, 0, 0,1, 0,1, 1, 1, 1, 1, 2, 0)$ & $0$ & $0$ & $0$ & $0$ & $0$& $^1\!\bar E_g \,^2\!\bar E_g\uparrow G|_{2c}$\\
11&$(2, 0, 0,1, 0,1, 1, 1, 1, 1, 0, 2)$ & $0$ & $0$ & $0$ & $0$ & $2$&-\\
12&$(2, 0, 0,1, 0,1, 1, 1, 1, 1, 1,1)$ & $1$ & $0$ & $0$ & $1$ & $3$&-\\
 \hline
 13&$(1,1, 1,0, 1,0, 1, 1, 1, 1, 2, 0)$ & $1$ & $0$ & $0$ & $0$ & $3$&-\\
14&$(1,1, 1,0 ,1,0, 1, 1, 1, 1, 0, 2)$ & $1$ & $0$ & $0$ & $0$ & $1$&-\\
15&$(1,1, 1,0, 1,0, 1, 1, 1, 1, 1,1)$ & $0$ & $0$ & $0$ & $1$ & $0$&-\\
 \hline
 16&$(1,1, 1,0, 0,1, 1, 1, 1, 1, 2, 0)$ & $1$ & $0$ & $0$ & $0$ & $1$&-\\
17&$(1,1, 1,0, 0,1, 1, 1, 1, 1, 0, 2)$ & $1$ & $0$ & $0$ & $0$ & $3$&-\\
18&$(1,1, 1,0, 0,1, 1, 1, 1, 1, 1,1)$ & $0$ & $0$ & $0$ & $1$ & $0$&-\\
 \hline
 
\end{tabular}
\caption{Physically irreducible representations  and topological invariants for isolated 4-band groups.}
\label{t1}
\end{table}
 
We now comment briefly on the method used to compute the SI in Table~\ref{t1}. The strong $Z_2$ index is given by the well known Fu-Kane formula~\cite{FK}
\beq\label{Z2}
Z_2=\frac{1}{2}\sum_{K\in\mathrm{TRIM}} n_K^- \;\; \mathrm{mod} \;\;\; 2,
\eeq
where $ n_K^-$ is the number of bands with parity $-$ at the TRIM point $K$. For SG55, the set of TRIM vectors coincides with the eight high symmetry points.
This formula is easily evaluated with the help of Table~\ref{t2}, which gives the number of parity eigenstates for the twelve physically irreducible representations. The weak topological index $Z_{2w,1}$ \hbox{($Z_{2w,2}$, $Z_{2w,2}$)} is obtained by restricting the sum in Eq.~(\ref{Z2}) to TRIMs with $k_x\!=\!\pi$ ($k_y\!=\!\pi$, $k_z\!=\!\pi$). Finally, the $Z_4$ index is given by the Fu-Kane-like formula~\cite{quant}
\beq\label{FK4}
Z_4=\frac{1}{4}\sum_{K\in\mathrm{TRIM}} (n_K^+-n_K^-) \mod  \; 4.
\eeq

 \begin{table}[t]
\begin{tabular}{| c|c ||c|c|| c|  c || c||c ||c||c|| c|  c |}
\hline
$\bar \Gamma_5(2)$ & $\bar \Gamma_6(2)$ & $\bar R_5\bar R_5(4)$ & $\bar R_6\bar R_6(4)$ & $\bar S_5\bar S_5(4)$ & $\bar S_6\bar S_6(4)$ & $\bar T_3\bar T_4(4)$ & $\bar U_3\bar U_4(4)$ & $\bar X_3\bar X_4(4)$ & $\bar Y_3\bar Y_4(4)$ & $\bar Z_5(2)$ & $\bar Z_6(2)$ \\
 \hline
 $2+$ &  $2-$ &  $4+$  & $4-$ & $4+$ &  $4-$ &  $2+2-$ & $2+2-$ & $2+2-$ & $2+2-$ & $2+$ &  $2-$\\
 \hline
\end{tabular}
\caption{Number of parity eigenstates for the physically irreducible representations at the high symmetry points.}
\label{t2}
\end{table}

Note  that $14$ out of the $18$ sets of bands in Table~\ref{t1} have SI and, as a consequence, their topology is necessarily  non-trivial. 
This gives a very rich phase space to explore. On the other hand,  the irrep content at the high symmetry points in cases 1, 5, 8 and 10
coincides with the  EBR indicated in the last column. The corresponding subsets of bands \textit{might} therefore be trivial according to our previous discussion. The EBRs for the complementary bands are obtained by changing the irrep label from $g$ to $u$.

\vskip.5cm
\ni{\bf PHASE DIAGRAM FOR THE TB MODEL}
\vskip.5cm
The $18$ cases in Table~\ref{t1} reflect the possible orderings of the band energies at the high symmetry points in the BZ. The hamiltonian can be diagonalized analytically at all the high symmetry points. Note that, regarding possible orderings,  points $T$, $U$, $X$ and $Y$ are irrelevant, as there is only one type of irrep in each case.  It turns out that only $7$   out of the $15$ parameters in the Hamiltonian given by Eqs. (1-2) in the Main Text enter the formulae for the energies at the relevant high symmetry point. More concretely, the spectra at the relevant high symmetry points can be written in terms of the five independent combinations 
\beq
\delta_0\;,\;\delta_\pm=\alpha+\beta\pm\gamma\;,\; \xi_\pm=\sqrt{(\lambda_{10}\pm\lambda_{13})^2+\zeta^2_{233}}
\eeq
according to Table~\ref{t3}. Then, as long as $Z_4$ is even,  it is easy to choose the parameters in such a way that the desired ordering is obtained and the valence and conduction bands are separated by a gap throughout the BZ. For $Z=1$ or $3$, the bands must cross at a high symmetry line or plane and the phase is semimetallic~\cite{quant}.

\begin{figure}[h]
\begin{center}
\includegraphics[angle=0,width=.8\linewidth]{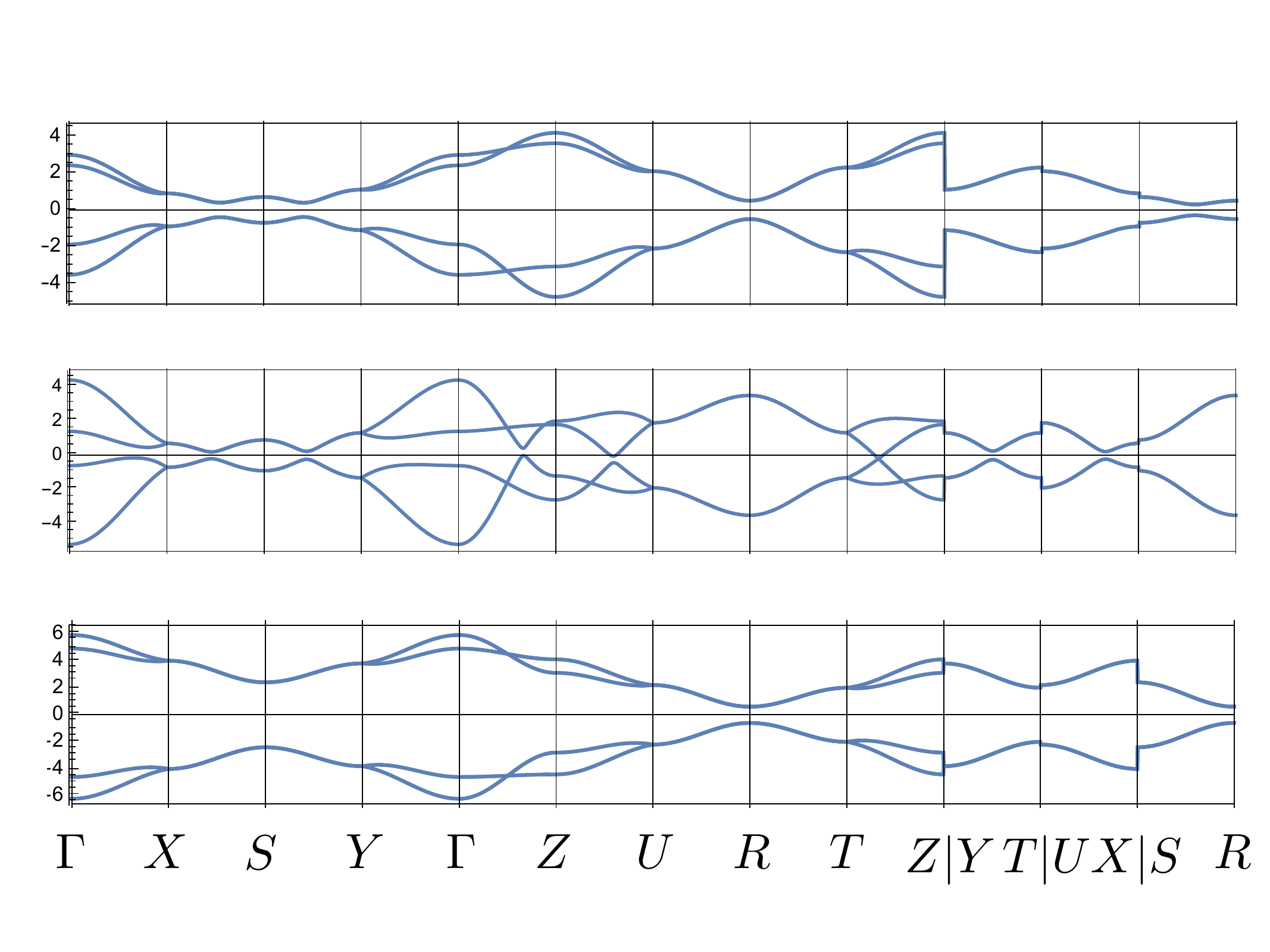}
\end{center}
\vspace*{-.8cm}
\caption{Bands for different values of the model parameters with energies in arbitrary units. Top: The only band inversion is at point $S$, $Z_4=2$, case $4$ in Table~\ref{t1}. Middle: Band inversions at the $R$, $S$ and $Z$ points, $Z_4=3$, case $12$. This is a semimetallic phase, note the band crossing at the $TZ$ line. Bottom: 
This phase has no SI and the irrep contents of the valence and conduction bands are those of EBRs, case $1$ in Table~\ref{t1}.}
\label{fig1}
\end{figure}
Three qualitatively different examples of bands for the TB model in the Main Text are presented in Fig.~\ref{fig1}. The top bands correspond to the phase studied in the main text, where there is an inversion only at point $S$. This is a TCI, with the irrep contents and SI given in case $4$ of Table~\ref{t1}. The middle bands correspond to case $12$ in Table~\ref{t1} and $Z_4=3$; there is a band crossing at the $TZ$ line and the phase is semimetallic. In the last example, at the bottom of Fig.~\ref{fig1}, all the SI vanish and the irrep contents of the valence and conduction bands coincide with the EBRs $^1\!\bar E_g \,^2\!\bar E_g\uparrow G|_{2a}$ and $^1\!\bar E_u \,^2\!\bar E_u\uparrow G|_{2a}$ respectively. There are no band inversions and the system is in an atomic limit.
\begin{table}[h]
\begin{tabular}{| c |c||c |c||c |c||c |c|}
\hline
$\bar \Gamma_5(2)$ & $\bar \Gamma_6(2)$ & $\bar R_5\bar R_5(4)$ & $\bar R_6\bar R_6(4)$ & $\bar S_5\bar S_5(4)$ & $\bar S_6\bar S_6(4)$ &  $\bar Z_5(2)$ & $\bar Z_6(2)$ \\
  \hline
 $ \delta_+ +\delta_0\pm\xi_+ $ &  $- \delta_+ -\delta_0\pm\xi_- $ & $-\delta_+ +\delta_0$& $\delta_+ -\delta_0$&$-\delta_- +\delta_0$&$\delta_- -\delta_0$& $ \delta_- +\delta_0 \pm \xi_+ $ &$- \delta_- -\delta_0\pm\xi_- $\\
  \hline
\end{tabular}
 \caption{Energies at the relevant  high symmetry points.}
\label{t3}
\end{table}
\\
\\




%

\end{document}